\documentclass[10pt,twoside]{article}
\usepackage{amssymb}
\usepackage{latexsym}
\usepackage{fullpage}
\usepackage{a4,latexsym,rotating,amsmath, amsfonts}
\usepackage{color}
\usepackage{graphicx}

\usepackage{amscd}
\usepackage{oldgerm}

\def\N{\mathbb{N}}
\def\Z{\mathbb{Z}}
\def\R{\mathbb{R}}

\newcommand{\grad}{\operatorname{grad}} 

\def\1{\mathbf{1}}
\def\:{\lrcorner}
\def\#{\sharp}

\def\d{\delta}
\def\e{\epsilon}
\def\r{\rho}
\def\o{\circ}
\def\V{\noindent}

\def\x{\otimes}
\def\<#1,#2>{\langle#1,\,#2\rangle}

\def\qed{\ensuremath{\quad\Box\quad}}
\def\pfill{\par\vskip2mm plus1mm minus1mm\noindent}
\def\inv#1{\raise.1em\hbox to 0pt{$^{-1}$\hss}_{#1}\;}

\def\v{\noindent}

\newcommand{\ov}{\overline}

\newcommand{\bean}{\begin{eqnarray*}}
\newcommand{\eean}{\end{eqnarray*}}
\newcommand{\benu}{\begin{enumerate}}
\newcommand{\eenu}{\end{enumerate}}
\newcommand{\eea}{\end{eqnarray}}
\newcommand{\bea}{\begin{eqnarray}}
\newtheorem{Theorem}{Theorem}
\newtheorem{Lemma}[Theorem]{Lemma}
\newtheorem{Corollary}[Theorem]{Corollary}
\newtheorem{Definition}[Theorem]{Definition}
\newtheorem{Example}[Theorem]{Example}

\newtheorem{Proposition}[Theorem]{Proposition}

\def\1{\mathbf{1}}
\def\:{\lrcorner}
\def\#{\sharp}

\def\x{\otimes}
\def\qed{\ensuremath{\quad\Box\quad}}
\def\pfill{\par\vskip2mm plus1mm minus1mm\noindent}
\def\inv#1{\raise.1em\hbox to 0pt{$^{-1}$\hss}_{#1}\;}

\def\v{\noindent}

\newcommand{\be}{\begin{equation}}
\newcommand{\ee}{\end{equation}}

\newcommand{\noi}{\noindent}
\newcommand{\sm}{\smallskip}

\newcommand{\ben}{\begin{enumerate}}
\newcommand{\een}{\end{enumerate}}
\newcommand{\bit}{\begin{itemize}}
\newcommand{\eit}{\end{itemize}}
\newcommand{\edoc}{\end{document}}

\newcommand{\bdefi}{\begin{Definition}}
\newcommand{\btheo}{\begin{Theorem}}
\newcommand{\bprop}{\begin{Proposition}}
\newcommand{\brema}{\begin{Remark}}
\newcommand{\bcoro}{\begin{Corollary}}
\newcommand{\blemm}{\begin{Lemma}}
\newcommand{\bexam}{\begin{Example}}

\newcommand{\edefi}{\end{Definition}}
\newcommand{\etheo}{\end{Theorem}}
\newcommand{\eprop}{\end{Proposition}}
\newcommand{\erema}{\end{Remark}}
\newcommand{\ecoro}{\end{Corollary}}
\newcommand{\elemm}{\end{Lemma}}
\newcommand{\eexam}{\end{Example}}

\newcommand{\cvd}{\ \rule{0.5em}{0.5em} \sm \noi}

\title{A note on invariant temporal functions}

\begin{document}

\author{Olaf M\"uller\footnote{Fakult\"at f\"ur Mathematik, Universit\"at Regensburg, D-93040 Regensburg, \texttt{Email: olaf.mueller@ur.de}}}

\maketitle

\begin{abstract}
\v The purpose of this article is to present a result on the existence of Cauchy temporal functions invariant by the action of 
a compact group of conformal transformations in arbitrary globally hyperbolic manifolds. 
Moreover, the previous results about the existence of Cauchy temporal functions 
with additional properties on arbitrary globally hyperbolic manifolds are unified in a very general theorem. 
To make the article more accessible for non-experts, 
and in the lack of an appropriate single reference for the Lorentzian geometry background of the result, 
the latter is provided in an introductory section.
\end{abstract}

\section{Introduction}

From the viewpoint of a global analyst, the appropriate class of geometries to consider for symmetric hyperbolic systems are globally hyperbolic manifolds. These are defined by assuming two causality conditions. Historically, the row of the most important results in this context is certainly the construction of continuous Cauchy time functions due to Geroch \cite{G} and of smooth Cauchy temporal functions due to Bernal and S\'anchez \cite{BS1}, \cite{BS2}. This latter statement ensures that in globally hyperbolic spacetimes, linear symmetric hyperbolic systems have well-defined initial value problems on Cauchy hypersurfaces. In a later work \cite{MS}, S\'anchez and the author proved a stronger statement ensuring even the existence of a {\em steep} Cauchy temporal function, thereby answering the question of an appropriate analogon of the Nash embedding theorem in the Lorentzian world. Now let us consider the initial value problem for Einstein theory, possibly coupled to any field theory. It is not an initial value problem of the sort mentioned above, as the metric is not a fixed background but a dynamical variable. Nevertheless, it is well-known by the work of Choquet-Bruhat and Geroch \cite{CG}, using Zorn's lemma, that constrained initial data give rise to a unique maximal solution, and the question of global existence is replaced by the question of geodesic completeness of the maximal solution. If the initial conditions are invariant under a group $G$ of diffeomorphisms (which are then isometries), the uniqueness statement implies that also the maximal solution is $G$-invariant. We can turn around the problem, begin with a $G$-invariant spacetime and ask if there are $G$-invariant initial conditions, that is, a $G$-invariant Cauchy surface and possibly even a $G$-invariant Cauchy temporal function. The author acknowledges Miguel S\'anchez for the proposal of this question. The present note answers it in the affirmative for the case of a compact group $G$. At the same time, it combines the previous existence result for steep Cauchy temporak functions of \cite{MS} with the problem, solved in \cite{BS3}, of finding a Cauchy temporal function adapted to a Cauchy surface $S$, that is, a Cauchy temporal function taking the value zero on $S$. The main result, Theorem \ref{main} below, states that we can require the three properties at once, steepness, adaptedness, and $G$-invariance for a compact group $G$ that, if we do not care about steepness, does not even need to consist of isometries, but is only assumed to consist of conformal diffeomorphisms. Applying Theorem \ref{main} in the case $m=0$ yields that we do not have to assume a priori invariance of any acausal subset under $G$, except for the case that we look for a steep Cauchy temporal function adapted to some Cauchy surface $S$. Then, of course, it is necessary to assume $G$-invariance of $S$. If we want to adapt the Cauchy temporal function to more than one Cauchy surface, we cannot require steepness. Explicitly, the theorem reads:

\begin{Theorem}
\label{main}
Let, for $n \in \N \cup \{\infty \}$, $ (M,g)$ be a $C^n$ Lorentzian manifold that is globally hyperbolic. Let $ k \leq n$, $m \in \N$ and\footnote{Throughout this article, we use the convention that $0 \in \N$} let $ (S^- = S_0, S_1, ...,  S_m, S^+ = S_{m+1})$ be an $(m+2)$-tuple of $C^k $ spacelike Cauchy hypersurfaces with $S_{i+1} \subset I^+(S_i)$ for all $0 \leq i \leq m+1$. Let $f_\pm: S^\pm \rightarrow \R$ be arbitrary continuous functions and let $a= (a_1, ... a_m)$ be an $m$-tuple of real numbers with $a_{i+1} > a_i$. Let $G$ be a compact group of time-oriented conformal diffeomorphisms of $(M,g)$. Let $t^\pm$ be future resp. past Cauchy time functions on $(I^+ (S^+),g)$ resp. $(I^-(S^-), g)$. Then there is a $C^{k-1}$ Cauchy temporal function $T$ with 
\begin{enumerate}
\item{$S_i= T^{-1} (\{ a_i \} )$ for all $i \in \{1, ... m \} $,}
\item{$ \pm T  >  \pm t^\pm /2   - 2$ on $I^\pm(S^\pm)$,}
\item{$\pm T \vert_{S^\pm} > f_\pm $.} 
\end{enumerate}
If $G$ leaves $S_i$ invariant for all $1 \leq i \leq m$, then $T$ can additionally been chosen $G$-invariant.  
If $m \in \{ 0, 1 \}†$ and if the group $G$ consists of isometries, we can moreover find such a $G$-invariant $T$ that is additionally steep. 
\end{Theorem}

\v The following proposition could easily be derived as a corollary of Theorem \ref{main}:

\begin{Proposition}
\label{AcausalOrbits}
 Let $G$ be a compact group of conformal time-oriented diffeomorphisms of a globally hyperbolic spacetime. Then each $G$-orbit is acausal.
\end{Proposition}

\v However, we will give an independent proof of the proposition at the beginning of the third section already before entering the constructions.

\bigskip
 
\v The article is structured as follows: 

\bigskip

\v The second section does not contain any new material. It is a short introduction into globally hyperbolic manifolds and continuous Cauchy time functions \`a la Geroch. All results can be found in greater detail in \cite{BEE}, \cite{BS4} and their references, but have been arranged in a way more focused on globally hyperbolic spacetimes. As the author thinks that such a short introduction could be helpful for beginners in the area, it has been written in a self-contained way and does not require prior knowledge of Lorentzian geometry.

\bigskip

\v Aim of the third section is the proof of Theorem \ref{main}. The first proposition in this section, Prop. \ref{t1}, and its proof is a slightly adapted version of the corresponding theorem and its proof in \cite{MS}. The theorem of this section comprises all previously published results on Cauchy temporal functions on globally hyperbolic manifolds, and even its two immediate corollaries, firstly the existence of a $G$-invariant smooth Cauchy temporal function  and secondly of a steep Cauchy temporal function adapted to a given Cauchy surface, are new results.

\section{Basics on Lorentzian manifolds and construction of continuous Cauchy time functions}

\V First of all, we have to do a bit of Lorentzian linear algebra, i.e. to consider bilinear forms of signature $(1,n-1)$ on an $n$-dimensional vector space:

\begin{Definition}
The {\bf signature} of a nondegenerate symmetric bilinear form $B$ on $\R^n$ is the tuple $(r,s)$ where $r$ is the number of negative and $s$ is the number of positive eigenvalues of the endomorphism $i_h^{-1} \o i_B$ where $h$ is a positive definite bilinear form (here, for a bilinear form $Z$, by $i_Z : \R^n \rightarrow (\R^n)^*$ we denote the insertion into $Z$ defined by $i_Z(v) := Z(v, \cdot)$). A vector $v \in \R^n$ is called {\bf spacelike} resp. {\bf timelike} resp. {\bf lightlike} iff $B(v,v) >0$ resp. $B(v,v) <0$ resp. $B(v,v) = 0$. 
\end{Definition}

\v We observe that the nonspacelike nonzero vectors form a double cone $C$ with tip $0$, having two connected components. If we choose a timelike vector $v$, then one connected component, say $C_1$, has the property that each vector $w$ in it satisfies $B(v,w) >0$, whereas each vector $u$ in the other component $C_2 = -C_1$ satisfies $ B(v,u) <0 $.

\begin{Theorem}
\begin{enumerate}
\item{The inverse Cauchy-Schwarz inequality holds: For all nonspacelike $v,w \in \R^n$ we have $  B(v,w) ^2 \geq B(v,v) \cdot B(w,w) $}
\item{The inverse triangle inequality holds: For every two nonspacelike vectors $v,w \in \R^n$ in the same connected component of nonspacelike vectors we have $\sqrt{-B(v+w, v+w)} \geq \sqrt{-B(v,v)} + \sqrt{-B(w,w)}$. Equality holds if and only if the vectors are lightlike and linearly dependent.}
\item{$B(v+w, v+w) \leq B(v,v)$ for all $v,w$ nonspacelike within the same connected component of $C$. Equality holds if and only if the vectors are linearly dependent.}
\item{Let $B, B'$ be bilinear forms on $\R^n$ with $B$ of signature $(1, n-1)$ and assume $B(v,v) = 0 \Rightarrow B'(v,v) $ for all $v \in \R^n$. Then there is $c \in R$ with $B'= cB$.}
\end{enumerate}
\end{Theorem}

\v {\bf Proof:} Straightforward. \hfill \cvd

\begin{Definition}
Let $M$ be an $n$-dimensional manifold, let $\tau M : TM \rightarrow M$ be the tangent bundle of $M$. A {\bf Lorentzian metric on $M$} is a symmetric section $g$ of $\tau^* M \x \tau^* M$ such that $g_p$ is of signature $(1,n-1)$ (via an arbitrary isomorphism $T_pM \rightarrow \R^n$).
\end{Definition}

\v {\bf Remark:} Unlike Riemannian metrics, Lorentzian metrics on a vector space do not form a convex cone and thus Lorentzian metrics need not to exist on an arbitrary finite-dimensional manifold. On compact manifolds, the Euler class is the only obstruction for the existence of a Lorentzian metric: Lorentzian metrics exist iff the Euler class vanishes. On a noncompact manifold there are always Lorentzian metrics. Keep in mind that the trace of bilinear forms $B$ in Pseudo-Riemannian vector spaces is different: $tr_g B := tr(i_g^{-1} \o i_B) = \sum \e_i B(e_i, e_i)$ for a pseudo-orthonormal basis $e_i$, e.g. for the Ricci tensor.

\bigskip

\v We consider $I_g:= \{¬†v \in TM \vert g(v,v) <0 \} $, the subset of timelike vectors, and $J_g:= \{¬†v \in TM \vert g(v,v) \leq 0 , v \neq 0 \} $, the subset of causal vectors. For $p \in M$, we denote $J_{g,p} = J_g \cap T_pM$ and $I_{g,p} := I_g \cap T_pM$. We observe that both $ I_{g,p} $ and $J_{g,p}$ have two components.

\begin{Definition}
A Lorentzian manifold $(M,g)$ is called {\bf time-orientable} or {\bf space-time} iff $J_g$ is disconnected.
\end{Definition}

\v If $(M,g)$ is time-oriented, then by standard arguments for sections into subsets of fiberwise nonempty interior, $J_g $ has exactly two components. Every choice of a connected component $J_g^+$ is called {\bf time-orientation of $(M,g)$} and $J_g^+$ is called future. The subset $J_g^- := - J_g^+$ is called past, and we have $J_g= J_g^+ \cup J_g^-$. 

\v We want to transfer this notion of 'future' from $TM$ to $M$, by means of curves.

\begin{Definition}   
A $C^0$ and piecewise $C^1$ curve\footnote{This regularity can be weakened, which is conceptually often more satisfying; for simplicity we nevertheless cling to piecewise $C^1$ curves in this introduction.} $c: I \rightarrow M$, where $I$ is a non-singleton interval, is called {\bf future-directed resp. past-directed causal curve} or short {\bf future resp. past curve} iff $\dot{c} (t) \in J^+_g$ resp. $\dot{c} (t) \in J^-_g$ for all $t \in I_j$ where the $I_j$ are the differentiability intervals of $c$. The set of future curves from $p$ to $q$ is denoted by $J_p^q$ and that of timelike future curves by $I_p^q$. Then we define $p \ll q $ iff $ I_p^q \neq \emptyset$, $p \leq q $ iff either $p=q$ or $ J_p^q \neq \emptyset $ and $J_g^\pm (p) := \{q \in M \vert p \leq q  \}$ and  $I_g^\pm (p) := \{q \in M \vert p \ll q  \}$. If we restrict to curves lying completely in a subset $U$ of $M$ we write instead $I_g^\pm(p, U) $ and $J_g^\pm (p,U)$. We also write, for a subset $A$ of $M$, $I^\pm (A) := \bigcup \{ I^\pm (p) \vert p \in A  \}$, and correspondingly for $J^\pm$.   
\end{Definition}

\begin{Theorem}
\label{OpenClosed}
Let $(M,g)$ be a spacetime.
\begin{enumerate}
\item{$I^+ (A)$ is open for $A$ an arbitrary subset.}
\item{{\bf Push-up property:} If $p \ll r \leq q$ or $p \leq r \ll q$, then $p \ll q$.} 
\item{$J^+(p) \subset \overline{I^+(p)}$.}
\end{enumerate}
\end{Theorem}

\v {\bf Proof.} 

\begin{Lemma}
\label{SafeNeighborhood}
For each point $p$ of a spacetime and each neighborhood $V$ of $p$ there is a {\bf $p$-safe neighborhood of $p$} contained in $V$, i.e., a geodesically convex neighborhood $U_p$ of $p$ contained in $V$ such that 

\begin{enumerate}
\item $I^\pm (p,U_p) \subset \exp_p (I^\pm_p)$, 
\item $I^\pm (p, U_p) $ open in $U_p$ (and thus in $M$),
\item $J^\pm (p, U_p) = \ov{I^\pm (p, U)} \cap U \subset \exp_p ( J^\pm_p)$  .
\end{enumerate}
\end{Lemma}

\v {\bf Proof of the lemma.} Take a convex neighborhood $U_0$ of $p$ contained in $V$ and let $U \subset U_0$ such that $\partial_0$ and $\grad x_0$ are still timelike. Then define $f: U \rightarrow \R$ by $f(q) := g_p (\exp_p^{-1} (q) , \exp_p^{-1} (q))$. Define $U^\pm := f^{-1} ((- \infty, 0)) \cap x_0^{-1} (\pm (0, \infty))$.  Gauss's lemma ($\exp$ is a radial isometry) implies that $\grad f$ is past timelike on $U^+ $ and future timelike on $U^-$, therefore any timelike curve $c$ from $p$ (which initially necessarliy enters $U^+$= has to stay in $U^+$ as long as it stays in $U$. Thus $I^+(p,U) \subset U^+ = \exp (I_p^+ )$. and the latter is open as $I_p^+$ is and $\exp$ is a diffeomorphism in $U$. As for the last statement, the inclusion "$\supset$" is obvious by continuity of $\exp$, the other one follows e.g. by stability arguments: Let $q \in J^+(p, U_p)$. If we slightly perturb the metric $g$ by $g_s:= g_0 + s da \x da  $ for a compactly supported one-form $a $ with $a \vert_U =  dx_0$, then for small $s$, say in $[0,1]$, $g_s$ is a spacetime, and $J^+_g \subset I^+_{g_s}$ for all $s>0$. Applying the second statement, that means that $r_n := ((\exp_p^{g_s}) \vert^U)^{-1} (q) \in I_{p, g_s}^+ $ for all $s \in [0,1]$. Now, as $ g_s (p) \rightarrow g $ w.r.t. a fixed Euclidean metric $e$ on $T_pM$, we have $ I^+_{g_s, p} \rightarrow I^+_{g,p} $, thus there are $r_n' \in I_{g,p}^+$ with $ \vert \vert r'_n - r_n \vert \vert_e \rightarrow 0$. We define $q_n := \exp_p^g (r_n') $, then $q_n \in I_g^+ (p) $. On the other hand, $\exp ^{g_s}$ converges to $\exp ^g$ uniformly in every compact subset of $T_pM$, thus by an easy $3 \e$-argument we have $q_n $ ßrightarrow $q$. \hfill \qed      

\bigskip

\v Now the proofs of all three statements of the theorem can be done easily by covering the curve in question by finitely many appropriately chosen safe neighborhoods as in the lemma and find an appropriate slightly perturbed timelike curve from the same initial point, and similar techniques. \hfill \cvd

\bigskip

\v Let $S,T$ be topological spaces and let $f $ be a map from $S$ into the set ${\rm Pot} (T)$ of subsets of $T$. Then $f$ is called {\bf inner continuous} if for all $s \in S$ and all compact subsets $ K \subset {\rm int} {f(s)}$ there is a neighborhood $U$ of $s$ in $S$ such that $K \subset {\rm int} f(u)$ for all $u \in U$. Analogously, $f$ is called {\bf outer continuous} if for all $s \in S$ and all compact subsets $ K \subset {\rm int} (M \setminus f(s)) $ there is a neighborhood $U$ of $s$ in $S$ such that $K \subset {\rm int} (M \setminus f(u)) $ for all $u \in U$.   

\begin{Theorem}
\label{InnerContinuous}
Let $(M,g)$ a spacetime, then the maps $I^\pm$ are inner continuous.
\end{Theorem}

\v {\bf Proof.} Let $p \in M$ and let $K \subset I^+(p)$ be compact. First we show that for any $q \in I^+(p)$, there are open neighborhhods $U_q$ of $q$ and $W_q$ of $p$ with $U_q \subset I^+(z) $ for all $z \in W_q$. Indeed we can choose a timelike curve $c:[0,1] \rightarrow M$ from $p$ to $q$, and for $r := c(1/2) $, openness of timelike cones as in Theorem \ref{OpenClosed} implies that there are open neighborhoods $U_q$ of $q$ and $W_q$ of $p$ with $U_q \subset I^+(r)$ and $W_q \subset I^-(r)$, those have the desired properties. Now compactness of a subset $K \subset I^+(p)$ implies that there is a finite number of sets $U_{q(1)}, ... U_{q(n)}$ covering $K$. Then $W:= \bigcap_{i=1}^n W_{q(i)}$, still an open neighborhood of $p$, satisfies $K \subset I^+(W)$ for all $w \in W$. \hfill \cvd

\bigskip

\v Let $(M,g)$ be a spacetime. A subset $A \subset M$ is called {\bf causally convex} if, for any causal curve $C$ in $M$, $c^{-1} (A)$ is a connected interval (possibly empty or degenerate, i.e. a point). A subset $A$ is called {\bf achronal} resp. {\bf acausal} iff there is no timelike curve between any two points of $A$, i.e. $I^+(p) \cap A = \emptyset $ for any $p \in A$, resp. if there is no future curve between any two points in $A$. A subset is called {\bf future set} resp. {\bf past set} if $A= I^+(A)$ resp. $A= I^-(A) $. For a subset $A \subset M$, the {\bf past} resp. {\bf future (causal) domain of dependence of $A$} is the set $D^-(A)$ resp $D^+(A)$ of points of $M$ such that every $C^0$-inextendible\footnote{If here and in the following we speak of $C^0$-inextendible future curves --- meaning that after reparametrizing them to a bounded interval $I$ there is no larger interval to which we could extend the curve as continuous map ---  the reader should nevertheless keep in mind that we always assume causal curves to be piecewise $C^1$.} future resp. past curve meets $A$. The {\bf domain of dependence of $A$} is $D(A) := D^-(A) \cup D^+(A)$. A subset $A$ is called {\bf Cauchy subset} if it is acausal and $D(A) = M$. In other words, $A$ is a Cauchy subset if it is met exactly once by every $C^0$-inextendible future curve.

\bigskip

\v Here we want to mention the physical interpretation of causal curves: Timelike future curves represent possible trajectories of massive particles, whereas massless particles travel along null future curves. This leads to causality requirements: In order to avoid obvious paradoxa in the context of time travels and under the assumption of a the free will, one wants, for example, to forbid closed causal curves:

\begin{Definition}
A space-time $(M,g)$ is called 

\begin{enumerate}
\item{{\bf causal} iff $J^+ (p) \cap J^-(p) = \{ p \} $ for all $p \in M$, i.e. iff there is no closed future curve in $(M,g)$,}
\item{{\bf diamond-compact} iff $J^+(p) \cap J^-(q)$ is compact for all $p,q \in M$,}
\item{{\bf globally hyperbolic} iff it is causal and diamond-compact.}
\end{enumerate}
\end{Definition}

The following examples for globally hyperbolicity are easy to verify: 

\begin{enumerate}
\item{Pseudo-Riemannian products $(\R \times N, -dt^2 + k)$ for $(N,k)$ complete are globally hyperbolic, in particular Minkowski spaces are globally hyperbolic.}
\item{If $g,h$ are time-oriented Lorentz metrics on $M$ with $g$ globally hyperbolic and $J_h \subset J_g$, then $h$ is globally hyperbolic as well. In particular a metric $h$ conformally related to a globally hyperbolic metric $g$ is globally hyperbolic.}
\item{Causally convex subsets of globally hyperbolic manifolds are g.h.}
\end{enumerate}

\V It is easy to see that if $A$ is acausal, then $\int (D(A))$ is causally convex and does not contain closed causal curves.

\v Causality is necessary for the existence of solutions to linear symmetric hyperbolic systems for initial values on closed spacelike hypersurfaces: To see this, consider the Lorentzian torus $T^{1,1}:= \R^{1,1}/ \Z^2$, the wave equation on $T^{1,1}$, then it is easy to find counterexamples to existence due to the timelike periodicity required by the geometry. On the other hand, diamond-compactness is necessary for uniqueness of the solutions: Consider a vertical strip $\R \times (-1, 1) $in $\R^{1,1}$, which is causal as an open subset of a causal spacetime, but again it is easy to find counterexamples for uniqueness of solutions to the wave equation, e.g. by considering initial values vanishing on $(-1,1)$. Given both properties, we have {\em well-posedness for natural initial value problems} with initial values on Cauchy hypersurfaces, a topic outside of the scope of this introductory section. Now we want to perform a necessary step in this business: to construct so-called time functions on a given spacetime. A function $t$ is called {\bf time function} iff $t \o c$ is strictly increasing for every future curve $c$. If there is a time function on $(M,g)$ then $(M,g)$ is obviously causal. A function $t$ is called {\bf future-Cauchy} iff for every upper $C^0$-inextendible future curve $c:[0,b) \rightarrow M$, the set $t \o c([0,b)) $ is unbounded from above, and {\bf past-Cauchy} iff for every upper $C^0$-inextendible past curve $c:[0,b) \rightarrow M$, the set $ t \o c([0,b)) $ is unbounded from below. Here upper $C^0$-inextendibility means that one cannot extend the curve as a continuous map to $[0,b]$. A function is called {\bf Cauchy} iff it is future-Cauchy and past-Cauchy, or equivalently, iff for every $C^0$-inextendible curve $c$, the map $t\o c$ is surjective onto $\R$. If a time function is also Cauchy, then every of its level sets is a Cauchy subset. The relevance of Cauchy time functions becomes clear in the next theorem:

\begin{Theorem}
\label{decomposition}
Let $(M,g)$ be a spacetime, let $k \in \N \cup \{ \infty \}$ and let $t$ be a $C^k$ Cauchy time function on $(M,g)$. Then $(M,g)$ is $C^k$-diffeomorphic to $\R \times N$, where $N$ is any level set of $t$.
\end{Theorem}

\v {\bf Proof.} Let $X$ be a timelike vector field defining the time orientation. Choose a complete Riemannian metric $H$ on $(M,g)$ and rescale $X$ to an $H$-unit vector field $Y:= (H(X,X))^{-1/2} \cdot X$. Then $Y$ is a complete timelike vector field, and its flow $F_Y$ defines the $C^k$ diffeomorphism $D$ via $ D(x) := (t(x), n (x)) $ where $n(x)$ is the unique point of $N$ met by the integral curve of $F_Y$ starting at $x$. \hfill \cvd  
 
\bigskip

\v As we want to construct the functions by means of future and past cones, we have to deduce some of their topological properties first: 

\begin{Theorem}
\label{DiamondThenClosed}
If $(M,g)$ is diamond-compact, then $J^\pm (p)$ is closed for every $p \in M$.  
\end{Theorem}
 
\v {\bf Proof.} Assume $r \in \overline{J^+(p)} \setminus J^+(p)$ and $q \in I^+(r)$. Then there is a sequence $(r_n)_{n \in \N}$ in $I^+(p)$ converging to $r$, thus there is $N \in \N$ with $r_n <q$ for all $ n>N  $ as $I^-(q)$ is open, so $(r_n)_{n>N}$ is contained in the compact set $J^+(p) \cap J^-(q)$, thus the sequence has an accumulation point $x$ in $J^+(p) \cap J^-(q)$. But then because of uniqueness of a limit we get $r = x \in J^+(p) \cap J^-(q)$, in contradiction to the assumption that $r \notin J^+(p)$. \hfill \cvd
 
 \bigskip

\v Now, together with inner continuity proven in Theorem \ref{InnerContinuous}, we can conclude that the timelike future cones depend continuously of their initial points: 

\begin{Theorem}[Geroch]
If $(M,g) $ is a spacetime with $J^\pm(q) $ closed for all $q$, then $I^\pm$ are outer continuous. 
\end{Theorem}

\v {\bf Proof.} we perform the proof for $I^-$. Assume $p \in M$, $K \subset M \setminus \overline{I^-(p)}$, assume that there is a sequence $(p_n)_{n \in \N}$ converging to $p$ with a corresponding sequence $r_n \in K \cap \overline{I^- (p_n)}$. Now compactness of $K$ allows us to single out a subsequence $(u_n)_{n \in \N}$ of $(r_n)_{n \in \N}$ that converges to a point $u \in K$. As $K $ is a subset of the open set $ M \setminus \overline{I^-(p)}$, we can find $s \in I^-(u) \cap (M \setminus \overline{I^-(p)})$. As $I^+(s)$ is open, for large $m$, we have $u_m \in I^+(s)$, thus $p_n \in I^+(s) \subset J^+(s)$, and the latter set is closed, thus $p \in J^+(s)$. But then, $s \in J^-(p) \subset \overline{I^-(p)}$, contradiction. \hfill \cvd

\bigskip

\v This is the initial point for the construction of continuous time functions:

\begin{Theorem}[Geroch]
\label{Geroch}
Let $(M,g)$ be globally hyperbolic, let $\mu$ be a volume form measure on $M$ with $\mu(M) < \infty$. Then the functions $t^{\pm} $ defined by $t^\pm (x) := \mp \mu (J^\pm (x))$ are continuous time functions.
\end{Theorem}

\v {\bf Proof.} We argue for $t^-$. Inner continuity of $I^-$, ensured by Theorem \ref{InnerContinuous}, implies lower semi-continuity of $t^-$: Let $p_n \rightarrow p$, $\e >0$, then we have to show that for large $n$ we have $t^-(p_n) > t^-(p) - \e$. Inner regularity of the volume form measure $\mu$ implies that there is a compact set $K$ in $I^-(p)$ with $\mu (K) > \mu(I^-(p)) - \e = t^-(p) - \e$. Thus inner continuity concludes the argument. 

\v Analogously, outer continuity of $I^-$ implies upper semi-continuity of $t^-$.

\v Now to show that $t^-$ is strictly increasing along every future curve, let $p\leq q \in M$. Causality of $M$ implies that $q \in M \setminus J^-(p)$ and that the latter subset is open, following Theorem \ref{DiamondThenClosed}. Now apply Lemma \ref{SafeNeighborhood} to find a safe neighborhood of $q$ in $M \setminus J^-(p)$. Now $I^-(q, U_q)$ is open in $M$ (thus of nonvanishing measure) and disjoint from $J^-(p)$, thus $t^-(q) > t^-(p)$.  \hfill \cvd
   
\bigskip

\v {\bf Remark:} Note that $t^\pm$ are still not Cauchy, as they are bounded, and in general there are not even functions $\Phi: \R \rightarrow \R $ such that $\Phi \o t^\pm$ are Cauchy: Consider $M:= \R^{1,1} \setminus J^+(0)$ (which is globally hyperbolic as a causally convex subset), irrespectively of the used volume form measure --- Consider the $C^0$-inextendible curves $c_a: (- \infty, a) \rightarrow M$, $c_a (s) := (a,s)$ and their pasts $P_a:= I^-(c_a((- \infty, a)))$. Then $P_0 \subset P_1$, and ${\rm int} (P_1 \setminus P_0) \neq \emptyset $. Thus ${\rm lim}_{t \rightarrow \infty} t^+ c_0 = \mu (P_0) < \mu (P_1) = {\rm lim}_{t \rightarrow \infty} t^+ c_1 $.  

\bigskip

\v $(M,g)$ is called {\bf strongly causal} iff for every point $p \in M$ and any open neighborhood $U$ of $p$ there is an open causally convex neighborhood $W \subset U$ of $p$.

\begin{Lemma}
\label{strongly-causal}
If $(M,g)$ is diamond-compact and causal, then every point in $M$ and every neighborhood $U$ of $p$ admits a geodesically convex and causally convex neighborhood $C \subset U$. In particular, $(M,g)$ is strongly causal. 
\end{Lemma}

\v {\bf Proof.} Take a geodesically convex neighborhood $C \subset U$ of $p$ and then, for $x \in C$, consider future and past cones $J^\pm_C (x)$ in $C$ ($(C,g \vert_C )$ as a Lorentzian manifold). Take a timelike curve $c: I \rightarrow C$ with $c(0) = p$. Let $t$ be a time function on $(M,g)$ whose existence is ensured by Theorem \ref{Geroch}. Define $S_a := t^{-1} (t(c(a)))$. Outer continuity of $I^\pm$ implies immediately that there is an $b>0$ such that $I^+(c(-b)) \cap S_b$ and $I^-(c(b)) \cap S_{-b}$ are contained in $C$. Then $W:= I_C^+(c(-b)) \cap I_C^-(c(b))$ is contained in $C \cap I^+(S_{-b}) \cap I^-(S_b)$, contains $p$, and any past curve $k$ leaving $W$ will not return before reaching $S_{-b}$. But after that, no return to $W$ is possible for $k$ as $t$ is a time function. Analogously for future curves.  \hfill \cvd

\bigskip

\v We further need a property of future curves that is often called 'non-imprisonment':

\begin{Theorem}
Let $(M,g)$ be strongly causal. Let $c: [0, b) \rightarrow M$ be a $C^0$-inextendible future curve. Then $c([0, b))$ is not precompact. 
\end{Theorem}

\v {\bf Proof.} Assume by contradiction that $c([0, b))$ is precompact. Choose $p_n := c(b-\frac{1}{n})$ and let $p$ be an accumulation point of $(p_n)_{n \in \N}$. We can choose a precompact geodesically convex neighborhood $U$ of $p$ so small that $\partial_0$ is timelike, and a causally convex neighborhood $W$ with $\overline{W} \subset U$ of $p$. Then causal convexity implies that for some $d \in [0,b)$ we have $c ([d,b)) \in W$. On the other hand, it is easy to see that any future curve in $ W$ is $C^0$-extendible with endpoint in $\overline{W}$: Parametrize the curve by the coordinate $x_0$, which is strictly monotonously decreasing but bounded, thus has a limit, and in the coordinates one sees that also the other coordinates have a limit. Thus $c$ is $C^0$-extendible, contradiction.    \hfill \cvd

\bigskip

\v Now let us return to the question of the existence of a continuous Cauchy time function. Above we have found $(M,g)$ where $ A \o  t^\pm$ is not Cauchy for any diffeomorphism $A: \R \rightarrow \R$. But somehow the limit of $t^+$ seems to be the right one for future curves and the limit of $t^-$ for past curves. Therefore we try to combine them in the new ansatz 
$$t:= {\rm ln} (-t^- /t^+) . $$

\begin{Theorem}[Geroch]
\label{Geroch}
Let $(M,g)$ be globally hyperbolic, then $t$ is a continuous Cauchy time function.
\end{Theorem}

\v {\bf Proof.} We have to show that if $c: (a,b] \rightarrow M$ is a $C^0$-inextendible future curve then $\lim_{s \rightarrow a} t (c(s)) = - \infty  $, the proof of the time-dual assertion being analogous. Taking into account inner regularity of $\mu$ we only need to show that for $K \subset M$ compact there is $s \in (a,b)$ with $K \cap I^-(s) = \emptyset$. Compactness of $K$ implies that $K \subset \bigcup_{i=1}^n I^+(q_i) $, thus in the proof we can replace $K$ by $I^+(q)$ for some $q$. Assume by contradiction that there is a sequence $k_n $ in $K \cap I^-(s(a+1/n))$. Then for an accumulation point $k$ of the sequence and a point $m< k$ there is an $N \in \N$ such that $s(a + 1/n) \in I^+(m) $ for all $n \geq N$ because $I^+(m)$ is open. But then $c(( a , a+ 1/N ]) \subset J^+(m) \cap J^- (c(b))$ and thus $c(( a , a+ 1/N ])$ is precompact, contradiction.  \hfill \cvd

\section{Proof of the main theorem}

\v The aim of this section is to prove Theorem \ref{main}, that is, a $G$-invariant steep temporal function (possibly adapted to a given Cauchy surface) for a given compact group of conformal diffeomorphisms $G$. One necessary condition for this to hold is obviously that the orbits of $G$ are acausal. This statement, formulated in Proposition \ref{AcausalOrbits} can be proven also directly:

\v {\bf Proof of Proposition \ref{AcausalOrbits}.} Assume that there are $p \in M$ and $gp \in J^+(p)$, then there is a future causal curve $c:[0,1] \rightarrow M$ from $p$ to $gp$. Then, as the group $G$ acts conformally and preserves the time orientation, the curve $k:[0, \infty) \rightarrow M$ defined by $k \vert_{[n,n+1]}= g^n \o c$ for any $n \in \N$, is a piecewise $C^1$ future causal curve. Compactness of $G$ implies that the subset $\{ g^n \vert n \in \N \}$  has the identity, and thus also $g$, as accumulation points, thus $k$ meets every neighborhood of $p$ again after a prescribed time, in contradiction to strong causality. \hfill \cvd

\bigskip

\v Let $(M,g)$ be a spacetime and let $A$ be a subset of $M$. A continuously differentiable function on $M$ is called {\bf temporal on $A$} iff it its gradient is past timelike on $A$, and {\bf temporal} iff it is temporal on $M$. The interest in temporal functions becomes clear in the light of Theorem \ref{decomposition} as, if a temporal function $t$ exists, it is automatically in the proof of Theorem \ref{decomposition}, one can take $X= {\rm grad} (t)$ and gets not only a differential, but even a metric decomposition: $(M,g)$ is then isometric to $(\R \times N, -f^2 dt^2 + {\rm pr}_2^* g_t) $ where $f \in C^{\infty} (M, (0, \infty))$ and $t \mapsto g_t$ is a smooth curve from $\R$ to the space of Riemannian metrics on $N$, equipped with the smooth compact-open topology, a Fr\'echet space topology on the space of symmetric bilinear forms in which the Riemannian metrics form an open cone. 

\bigskip

\v Now, in order to improve this decomposition a bit, one can try to find a decomposition with $f$ as above bounded. That leads immediately to the notion of {\em steep} Cauchy temporal functions. A continuously differentiable function $t$ on $M$ is called {\bf steep on $A$} if there is some $c>0$ such that $g({\rm grad} (t), {\rm grad} (t) ) < - c^2$ on $A$. It is called {\bf steep} iff it is steep on $(M,g)$. Obviously, every steep function is temporal. The first result of this section, for whose proof we slightly adapt the proof given in \cite{MS}, is the existence of steep Cauchy functions:

\begin{Proposition}[Existence of steep Cauchy temporal functions, compare with \cite{MS}]
\label{t1}
Let $(M,g)$ be globally hyperbolic and let $t$ be a $C^0 $ Cauchy time function on $(M,g)$. Then $(M,g)$ admits a steep
Cauchy temporal function $t_1$ with $\vert t_1 \vert > \vert t \vert /2 - 1$ and, thus, is isometric to a manifold $(\R \times N, - f^2 dt_1^2 + g_t) $ where $f \in C^{\infty} (M, (0,1])$ and $r \mapsto g_r$ is a smooth curve in the space of Riemannian metrics on $N$, such that all level sets of $t_1$ are Cauchy surfaces.
\end{Proposition}

\v So, in what follows $(M,g)$ will be a globally hyperbolic
spacetime, and we will assume that $t$ is a Cauchy time function
as given by Theorem \ref{Geroch}. We also use the following notation:

$$
T^b_a= t^{-1}([a,b]), \quad \quad S_a= t^{-1}(a).$$

\v For a Cauchy surface $S$, we say that $p \in M$ is {\bf $S$-safe} 
iff there exists a geodesically convex neighborhood $U_p \subset V$ with
$\partial^+U_p \subset J^+(S) $, where
$\partial^+U_p:= \partial U_p \cap J^+(p)$.

\v In this case, let $j_p \in C^\infty (I^-(S)) $ be the function
$$q\mapsto j_p(q) =\exp({-1/\eta(p,q)}),$$

\v if $q \in I^+(p) \cap I^-(S) $ and $0$ otherwise, where $\eta(p,q) := \langle v , v \rangle$ for $exp_p(v) = q$ if $v$ is future timelike and $0$ otherwise. 
 For any two subsets $A,B\subset M$, we put

$$J(A,B):=J^+(A)\cap J^-(B)$$  
and $J(p,B):=J (\{ p\}, B¬†)$.

\blemm  \label{lemilla0} Let $\tau$ be a function  such that
$g(\nabla \tau, \nabla \tau) <0$ in some open subset $U$ and let
$K\subset U$ compact. For any function $f$ there exists a constant
$c$ such that $g(\nabla (f+c\tau), \nabla (f+c\tau)) <-1$ on $K$.

\elemm

\v {\bf Proof.} Notice that at each $x$ in the compact subset $K$ the
quadratic polynomial $g(\nabla (f(x)+c\tau(x)), \nabla
(f(x)+c\tau(x)))$ becomes smaller than -1
for some large $c$
\cvd

\blemm  \label{lemilla} Let $t$ be a Cauchy time function and let $S$ be a level set of $t$, let $p\in
J^-(S)$. For all neighborhood $V$ of $J(p,S)$ there exists a
smooth function $\tau\geq 0$ such that:

(i) $\hbox{{\rm supp}} \, \tau\subset V$

(ii) $\tau > 1$ on $S\cap J^+(p)$.

(iii) $\nabla \tau$ is timelike and past-directed in $\hbox{{\rm
Int(Supp} } (\tau) \cap J^-(S))$.

(iv) $g(\nabla \tau, \nabla \tau) <-1$ on $J(p,S)$.

\elemm

\v {\bf Proof.} Let $ t$ take the value $a$ on $S$, and let $K\subset V$ be
a compact subset such that $J(p,S_a) \subset $ Int $(K)$. This
compactness yields some $\delta >0$ such that for every $x\in K$, 
$x$ is $S_{t(x) + 2 \d}$-safe. 

\v Now, choose $a_0<a_1:=
t(p) < \dots < a_n = a$ with $a_{i+1}-a_i<\delta /2$, and
construct $\tau$ by induction on $n$ as follows.

\v For $n=1$, cover $J(p,S)=\{p\}$ with a set type $I^+(x)\cap U_x$
with $x\in K\cap T_{a_0}^{a_1}$ and consider the corresponding
function $j_x$. For a suitable constant $c>0$, the product $c j_x$
satisfies both, (ii), (iii) and (iv). To obtain smoothability
preserving (i), consider the open covering $\{I^-(S_{a+\delta}),
I^+(S_{a+\delta/2})\}$ of $M$, and the first function $0\leq
\mu\leq 1$ of the associated partition of the unity (Supp $\mu
\subset I^-(S_{a+\delta})$). The required function is just $\tau =
c\mu j_x$.

\v Now, assume by induction that the result follows for any chain
$a_0<\dots < a_{n-1}$. So, for any $k\leq n-1$, consider
$J(p,S_{a_k})$ and choose a compact set $\hat K \subset $ Int $K$
with $J(p,S)\subset $ Int $\hat K$. Then, there exists  a function
$\hat \tau$ which satisfies condition (i) above for $V=$ Int $\hat
K \cap I^-(S_{a_{k+1}})$ and conditions (ii), (iii), (iv) for
$S=S_{a_k}$. Now, cover $\hat K \cap T^{a_{k+1}}_{a_k}$ with a
finite number of sets type $I^+(x^i)\cap U_{x^i}$ with $x^i\in K
\cap T^{a_{k+1}}_{a_{k-1}}$, and consider the corresponding
functions $j_{x^i}$.

\v For a suitable constant $c>0$, the sum $\hat \tau + c \sum_i
j_{x^i}$ satisfies  (iii) for $S=S_{a_{k+1}}$. This is obvious in
$J^-(S_{a_k})$ (for any $c>0$), because of the convexity of
timelike cones and the reversed triangle inequality. To realize
that this can be also obtained in $T^{a_{k+1}}_{a_k}$, where $
\nabla \tau$ may be non-timelike, notice that the support of
$\nabla \hat\tau |_{T^{a_{k+1}}_{a_k}}$ is compact, and it is
included in the interior of the support of $\sum_i j_{x^i}$, where
the gradient of the sum is timelike, thus we can use Lemma \ref{lemilla0}.

As $J^+(p,S_{a_{k+1}})$ is compact, conditions  (ii), (iv) can be obtained by choosing, if necessary, a bigger $c$.

\v Finally, smoothability (and (i)), can be obtained again by using
the open covering $\{I^-(S_{a_{k+1}+\delta})$, $
I^+(S_{a_{k+1}+\delta/2})\}$ of $M$, and the corresponding first
function $\mu$ of the associated partition of the unity, i.e.
$\tau = \mu (\hat \tau + c \sum_i j_{x^i})$. \hfill \cvd

\bigskip

\v In order to extend  locally defined temporal functions to a global
temporal function, one cannot use a partition of the unity (as stressed in
previous proof, as $\nabla \tau$ is not always timelike when $\mu$
is non-constant). Instead,  local temporal functions must be added
directly. To that purpose, adapted coverings are needed as those provided in the following definition:

\bdefi Let $S$ be a Cauchy hypersurface. A {\em fat cone covering}
of $S$ is a sequence of points $p'_i\ll p_i , i\in \N$ such that
both, ${\cal C}'=\{I^+(p'_i): i\in \N\}$ and ${\cal
C}=\{I^+(p_i): i\in \N\}$ yield a locally finite covering of $S$.
 \edefi

 \blemm \label{fatcone} Any Cauchy hypersurface $S$ admits a fat cone covering $p'_i \ll p_i, i\in \N$.
\v Moreover, both ${\cal C}$ and ${\cal C}'$ yield also a finite
 subcovering of $J^+(S)$.
\elemm

\v {\bf Proof.} Let $\{K_j\}_j$ be a sequence of compact subsets of
$S$ satisfying $K_j\subset $ Int $K_{j+1}$, $S=\cup_jK_j$. Each
$K_j\backslash$ Int $K_{j-1}$ can be covered by a finite number of
sets type $I^+(p_{jk}), k=1\dots k_j$ such that $I^+(p_{jk})\cap S
\subset K_{j+1}\backslash K_{j-2}$. Moreover, by continuity of the set-valued function $I^+$, this last inclusion
is fulfilled if each $p_{jk}$ is replaced by some close
$p'_{jk}\ll p_{jk}$, and  the required pairs $p'_i (=p'_{jk})$,
$p_i (=p_{jk})$, are obtained.

\v For the last assertion, take $q\in J^+(S)$ and any compact
neighborhood $W \ni q$. As $J^-(W)\cap S$ is compact, it is
intersected only by finitely many elements of ${\cal C}, {\cal C}'$,
and the result follows. \hfill \cvd

\begin{Definition}
\label{fls} Let $p' , p \in T_{a-1}^a$, $ p'\ll p$. A {\em steep
forward cone  function} for $(a, p' , p) $ is a smooth
function $ h_{a,p',p}^+ : M \rightarrow [ 0 , \infty )$ which
satisfies the following:

\begin{enumerate}
\item{$supp (h_{a,p',p} ^+)  \subset J^+ (p', S_{a+2})$,}
 \item{$h_{a,p',p}^+ > 1$ on $
S_{a+1}\cap J^+ (p)$},
 \item{If $x
\in J^- (S_{a+1}) $ and $h_{a,p',p}^+ (x) \neq 0 $ then $ \nabla
h_{a,p',p} ^+ (x)$ is timelike and past-directed, and} \item{ $g(
\nabla h_{a,p',p}^+ , \nabla h_{a,p',p} ^+ ) < - 1 $ on
$J(p,S_{a+1})$.}
\end{enumerate}

\end{Definition}

\v Now, Lemma \ref{lemilla} applied to $S=S_{a+1},
V=I^-(S_{a+2})\cap I^+(p')$ yields directly:

\blemm For all $(a,p',p)$ there exists a steep forward cone function. \hfill \cvd \elemm

\v The existence of a fat cone covering (Proposition \ref{fatcone})
allows to find a function $h^a_+$ which in some sense globalizes
the properties of a steep forward cone function.

\blemm \label{l07} Choose  $a\in \R$ and take any fat cone
covering $ \{ p'_i\ll p_i \vert i\in \N \}$ for $S=S_a$. For every positive
sequence $\{ c_i \geq 1 \vert i \in \N \}$, the non-negative
function $h_{a}^+:=(|a|+1) \sum_i c_i h_{a,p'_i,p_i} ^+$
satisfies:

\begin{enumerate}
\item{$supp (h_{a} ^+)  \subset J(S_{a-1},S_{a+2})$,}
 \item{$h_{a}^+ > |a|+1$ on
 $S_{a+1}$},
 \item{If $x
\in J^- (S_{a+1}) $ and $h_{a}^+ (x) \neq 0 $ then $ \nabla h_{a}
^+ (x)$ is timelike and past-directed, and} \item{ $g( \nabla
h_{a}^+ , \nabla h_{a} ^+ ) < - 1 $ on $J(S_a, S_{a+1})$.}
\end{enumerate}
\elemm 

\v {\bf Proof:} Obvious. \hfill \cvd

\bigskip

\v The gradient of $h_a^+$ will be spacelike  at some subset of
$J(S_{a+1}, S_{a+2})$. So, in order to carry out the inductive
process which proves Theorem \ref{t2}, a strengthening of Lemma
\ref{l07} will be needed.

\blemm \label{l07b} Let $h^+_a\geq 0$ as in Lemma \ref{l07}. Then
there exists a function $h^+_{a+1}$ which satisfies all the
properties
 corresponding to Lemma \ref{l07} and additionally:

\be \label{eee} g(\nabla (h^+_{a}+  h^+_{a+1}), \nabla (h^+_{a}+
h^+_{a+1}))< -1 \quad \quad \hbox{on} \; J(S_{a+1}, S_{a+2}) \ee
 (so, this inequality holds automatically on all $ J(S_{a}, S_{a+2})$).
\elemm

\v {\bf Proof.} Take a fat cone covering $ \{ p'_i\ll p_i \vert i\in \N \}$ for $S=S_{a+1}$.
Now, for each $p_i$ consider a constant $c_i\geq 1$ such that $c_i
h_{a+1,p'_i,p_i} ^+ + h^+_a$ satisfies inequality (\ref{eee}) on
$J^+(p_i, S_{a+2})$ (see Lemma \ref{lemilla0}). The required
function is then 

$h_{a+1}^+=(|a|+2) \sum_i c_i  h_{a+1,p'_i,p_i}
^+$.  \hfill \cvd

\bigskip

\v Now, we have the elements to complete the proof of Proposition \ref{t1}.

\bigskip

\noi {\bf Proof of Proposition \ref{t1}.} Consider the function
$h^+_a$ provided by Lemma \ref{l07} for $a=0$, and apply
inductively Lemma \ref{l07b} for $a=n\in \N$. Then, we obtain a
function $t_1^+ = \sum_{n=0}^\infty h_n^+ \geq 0$ with nowhere
spacelike gradient, which is a steep temporal function on
$J^+(S_0)$ with support in $J^+(S_{-1})$. Analogously, one can
obtain a function $t_1^-\geq 0$ which is a steep temporal
function with the reversed time orientation, on $J^-(S_0)$. So,
$t_1 := t_1^+- t_1^-$ is clearly a steep temporal
function on all $M$.

\v Moreover, the levels hypersurfaces of $t_1$ are Cauchy. In
fact, consider  any future-directed causal curve  $\gamma$, and
reparametrize it with the Cauchy time function $t_0$. Then, given a point $x$ with $t(x) = a \in (0, \infty )$, by using the Gau\ss \  bracket $[\cdot ]$ we obtain for all $x \in J^-(S_{-1}) \cup J^+(S_0)$:

$$ t_1 (x) \geq {\rm inf } t_1 (t^{-1} ([a])) = {\rm inf} t_1^+( t^{-1} ([a])) \geq {\rm inf} h^+_n( t^{-1} ([a])) > [a] +1 \geq a = t(x) $$ 

\v and a corresponding statement for $ a \in ( - \infty, 0) $. Thus $\vert t_1 \vert > \vert t \vert /2$ on $J^-(S_{-1}) \cup J^+(S_0)$.

\v In particular, any $C^0$-inextendible future curve $\gamma$ crosses all the levels of $t_1$, as required. \hfill \cvd

\bigskip

\v We call a $C^1$ function $s$ on a spacetime $(M,g)$ {\bf almost-temporal} iff $\nabla s $ is zero or past everywhere.

\begin{Lemma}
\label{t2}
Let $(M,g)$ be globally hyperbolic, let $t$ be a $C^0$ Cauchy time function on $(M,g)$ and let $S^-$ and $S^+$ be Cauchy surfaces of $(M,g)$ with $S^+ \subset I^+(S^-)$. Let $f^\pm$ be a continuous positive function on $S^\pm$. Then there are steep Cauchy temporal functions $t_2^\pm( S^\pm, t)$ on $(M,g)$ with: 
\begin{enumerate}
\item{$  t_2^+ ( S^+, t) >  t  /2 - 1$ on $ J^+(S^+) $ and $ t_2^+ (S^+, t) \vert_{S^+} >  f^+ $,}
\item{$  t_2^- (S^-, t) <  t  /2 + 1$ on $J^- (S^-)$ and $ t_2^- (  S^-, t) \vert_{S^-} <  f^-$.}
\end{enumerate}
\end{Lemma}

\v {\bf Proof.} First we choose an intermediate Cauchy surface $S$ with $S^\pm \subset I^\pm (S)$. For the first assertion, take a steep Cauchy temporal function $t_1$ on $(M,g)$ with $ \vert t_1 \vert > \vert t \vert /2 - 1$ on $J^-(S) \cup J^+(S^+)$ as in Proposition \ref{t1}. Then choose a fat cone covering $\{ (p_i', p_i ) \} $ of $S^+$. For each $i$ we choose a steep temporal function $t^i$ on $I^+(p'_i)$. Now, as $C_i:= J^+(p_i)  \cap S^+$ is a compact subset of $I^+(p'_i)$, $t^i$ takes a minimum $m_i$ on $C_i $. Now take any smooth increasing function $\phi_i: \R \rightarrow \R$ with $\phi_i ((- \infty, m_i-1)) = \{ 0 \} $ and $ \phi_i' (x) = 1 \forall x \geq m_i$. Then $s_i:= \phi_i \o t^i$, extended by $0$ on $M \setminus I^+(p'_i)$ is a smooth nonnegative almost-temporal function which is steep on $I^+(C_i)$. By choosing an appropriate constant $c_i$ one can ensure that $t_1 + c s_i > f$ on $C_i$. Now we consider  $t_2^+ (S,t):= t_1 + \sum_i c_i s_i$. Recall that the sum is a well-defined smooth function as the $I^+(p'_i) \cap S$ have been chosen locally finite. And $  t_2 (S^+, t) >  t  /2 - 1$ continues to hold on $J^+( \{ t>0 \} )† $ for any choice of constants as we only add nonnegative terms (to show this, it is useful to distinguish the cases $t<0$ and $t \geq 0$). For the second assertion, just time-dualize. \hfill \cvd

\begin{Lemma}
\label{t3}
Let $ (M,g)$ be globally hyperbolic, let $S^- < S < S^+$ be $C^k $ Cauchy hypersurfaces, let $S$ be spacelike and of regularity $C^k$. Let $t^\pm$ be Cauchy time functions on $I^\pm(S)$. Then there is a $C^{k-1}$ steep Cauchy temporal function $t_3$ with $t_3 (S) =0$, $\pm t_3 \vert_{S^\pm} > f^\pm $ and $  \pm t_3  > \pm  t^\pm /2 - 2$ on $I^\pm(S)$.
\end{Lemma}

\v {\bf Proof.} Let us first, along the lines of the corresponding proof in \cite{BS3}, construct a temporal function around $S$, which will then turn out to be steep on $S$. 

\v Let $\nu: S \rightarrow TM$ be the future normal vector field of $S$ and let $W$ be a normal neighborhood of $S$, i.e., we assume that $E: S \times \R \supset A \rightarrow W$ is a diffeomorphism, where $ E(s,r) := \exp (r \cdot \nu (s))$, and where $A \cap (\{ s \} \times \R)$ is a connected open interval containing $0$. We define a causally convex subneighborhood $U \subset W$ of $S$ in the following way: Choose a covering of $S$ by subsets $I^-(p_i)$ for $p_i \in I^+(S)$ with $d(p_i, S) <1$ and such that $U_i := I^-(p_i) \cap J^+(S) \subset W$. This can be constructed by considering Pseudo-Riemannian normal coordinates in a causally convex and geodesically convex neighborhood contained in $W$ of a point $p \in S$. Analogously choose a covering of $S$ by subsets $I^+(q_j)$ for $q_j \in I^-(S)$ such that $V_j := I^+(q_j) \cap J^-(S) \subset W$. Then the union $U$ of $\bigcup_{i \in I} U_i $ and $\bigcup_{j \in J} V_j$ is easily seen to be causally convex and contained in $W$. Now we want to produce an almost-temporal function out of the signed distance function $\d_S$ of $S$, which is $C^{k-1}$ in $W$. To that aim, observe first that $U^\pm := U \cap I^\pm(S)$ is also causally convex, thus globally hyperbolic and therefore has a Cauchy surface $S^{++}$ resp. $S^{--}$ which is also a Cauchy surface for $I^\pm (S)$. Let $V := I^+(S^{--}) \cap I^-(S^{++})$.

\v Now we construct two almost-temporal functions $\theta^+: M \rightarrow [0,1]$ and $\theta^-: M \rightarrow [-1, 0]$ with the properties:

\begin{enumerate}
\item{$\theta^+ (J^-(S^{--}) ) = \{†0 \}†$, $\theta^+ (J^+(S)) = \{ 1 \}$,}
\item{$\theta^- (J^-(S) ) = \{-1 \}$, $\theta^- (J^+(S^{++})) = \{ 0 \}$.}
\end{enumerate}

\v These functions can be defined easily by appropriate reparametrizations of Cauchy temporal functions on $V^\pm$. Then we put

$$\theta :=   2 \frac{(\d_s + 1) \theta^+}{(\d_s + 1) \theta^+ - \theta^-} - 1 , $$

\v and we observe easily that $\theta$ is almost-temporal, $ \theta (S) = 0$, $\theta$ can be smoothly extended to all of $M$ by $1$ on $J^+(S^{++})$ and by $-1$ on $J^-(S^{--})$. Moreover, 

$$ \nabla \theta \vert_S = \nabla \d_s \vert_S , $$

\v thus $\theta$ is steep on $S$. As $\theta$ is a $C^1$ function, $\theta$ is steep in an open neighborhood $V_0$ of $S$. As each open neighborhood of $S $ contains a causally convex subneighborhood of $S$ (see above) we can find Cauchy surfaces $\tilde{S}^\pm \subset I^\pm (S) \cap V_0$. Now apply Lemma \ref{t2} to $(I^\pm (S), \tilde{S}^\pm)$ to get steep Cauchy temporal functions $t^{\pm}$ on $I^\pm (S)$ with $\pm t^\pm \vert_{\tilde{S}^\pm}>  1$. Then choose smooth functions $\phi^\pm: \r \rightarrow \R$ with $\phi^+$ increasing, $\phi^+(x) = 0 $ for all $ x <0$, $\phi^+ (x) \geq x $ for all $x \geq 1$ and $(\phi^+)'(x) \geq 1 $ for all $x \geq 1$ and $\phi^- (y) = - \phi^+ (-y)$ for all $y \in \R$. Then define functions $s^\pm$ on $M$ by $s^+ (x) := \phi^+ ( t^+ (x)) $ for all $x \in I^+(S)$ and $s^+(x) := 0$ for all $ x \in J^-(S)$, as well as $ s^-(x) := - \phi^- (- t^- (x)) $ for all $x \in I^-(S)$ and $s^-(x) := 0$ for all $ x \in J^+(S)$. Now, obviously, the function

$$ \tilde{t}_3:= s^- + \theta + s^+$$

\v is steep on all of $M$, vanishes on $S$ and satisfies the requirement involving $t^{\pm}$. To satisfy the requirement involving $f^\pm$ as well, we find another pair of smooth Cauchy temporal functions $T^\pm$ on $I^\pm (S) $ as in Lemma \ref{t2} with $  T^+ \vert_{S^+} > \max \{  f^+, 1 \} $ and $T^- \vert_{S^-}< \min \{ f^-, -1 \}$ and set $Z^\pm := \phi^\pm \o T^\pm $ as in te definition of the $s^\pm$  above and

\bean
t_3 := Z^-  + \tilde{t}_3 + Z^+ ,
\eean

\V which finally satisfies all our requirements.  \hfill \cvd

\bigskip

\v Note that, given an additional $G$-symmetry fixing $S$, one possible procedure in the proof above would have been to adapt $U$ to the group action by setting $U := \bigcap_{g \in G} g(U)$. It can be shown that due to compactness of $G$ this is still a nonempty open neighborhood of $S$, and it is still causally convex as intersection of causally convex subsets. However, we will not need this construction in the following.

\v Finally, we have all elements at hand for the proof of the main theorem that takes into account additionally a given compact group $G$ of conformal diffeomorphisms:

\bigskip

\v {\bf Proof of Theorem \ref{main}.} 
Let $\omega$ be a volume form on $M$. Then we can construct an invariant volume form $\omega^G$ on $M$ by the well-known averaging process via the left-invariant Haar measure $\mu$ on $G$: 

\bean
\omega^G := \int_G g^* \omega dg .
\eean

\V By convexity of $\Omega^n M$ it is easy to see that $\omega^G$ is indeed a smooth top form, and the normalization of $ \mu$ entails that $\omega^G $ is indeed a volume form, which is, moreover, obviously left-invariant under $G$. Let $S_1$ be $G$-invariant and let $t^\pm$ be the Geroch time functions for the volume form $\omega^G$. As $G$ consists of conformal diffeomorphisms, 

\bean
g (I^\pm (p)) = I^\pm (g(p)) ,
\eean

\V for all $p \in M$, and this, together with the $G$-invariance of $S_1$ and $\omega^G$, implies that $t^\pm $ are $G$-invariant time functions on $I^\pm (S_1)$. Now let us first assume $m=1$, and let $S:= S_1$. Let $t_3$ be a smooth steep Cauchy temporal function with $t_3 (S) =0$ and $ \pm t_3^\pm  >  t^\pm /2 \pm 2$ on $I^\pm (S^\pm)$. Existence of $t_3$ is ensured by Lemma \ref{t3}. Then 

\bean
t_4 := \int_G g^*  t_3 d \mu (g) 
\eean

\V is a $G$-invariant smooth function, well-defined because of compactness of $G$ and the usual estimates. As $S_A$ is $G$-invariant, it is adapted to $S$. If $G$ consists of isometries, it is steep temporal because each map $dg$ preserves $ \{v \in TM \vert g(v,v) < a \} $ which is a fiberwise convex subset due to the inverse triangle inequality in $T_pM$, and because 

\bean
d t_4 (x) = \int_G d (g^* t_3) d \mu (g) = \int_G (dg \o dt_3) d \mu(g) ,  
\eean

\V as we can commute derivative and integral due to compactness of $G$. Finally, it is Cauchy as, given a number $r \in \R$, every $C^0$-inextendible future curve $c$ has to reach $(t^+)^{-1} (2r) $, and $t_4 (x) \geq r-1$ for every $x \in (t^+)^{-1} (2r) $, a property inherited by the function $t_3$ (as all level sets of $t^+$ are $G$-invariant). The corresponding statement holds for $C^0$-inextendible past curves. Thus $t_4 \o c$ is surjective onto $\R$ for $C^0$-inextendible causal curves. The statement on a finite sequence of Cauchy surfaces, i.e. on the case $m>1$, follows by an easy inductive argument:

\V Let $a_i, S_i$ be given for all $i \in \{ 1, ... , n+1 \} $. Then by inductional assumption there is $t^{[(1, ... , n)]}$ $G$-invariant with $t^{[(1, ... , n)]} (S_i) = \{ a_i\} $ and $t^{[(1, ... , n)]} \vert_{S_{n+1}} > a_n $.  We set 

\bean
t^{[(1, ... , n+1)]} := \phi \o t^{[(1, ... , n)]} + \psi \o t^{[n+1]}, 
\eean 

\V where $t^{[n+1]}$ is $G$-invariant, $t^{[n+1]} (S_{n+1}) = \{ \frac{a_{n+1}- a_n}{2} \} $, $\psi, \phi: \R \rightarrow \R$ are monotonously increasing, $\phi (x) = x $ for all $x \leq a_n$, $\phi (x) = \frac{a_n + a_{n+1}}{2}$ for all $ x \geq \frac{a_n + a_{n+1}}{2}$, and $\psi (x) = 0$ for all $x \leq 0$ and $ \psi (x) = x $ for all $x \geq \frac{a_{n+1}- a_n}{2}$.

 \hfill \cvd


\begin{thebibliography}{99}

\bibitem{BEE}
John K. Beem, Paul Ehrlich, Kevin Easley: {\em Global Lorentzian geometry}, 2nd edition. CRC Press (1996)

\bibitem{BS1}
Antonio Bernal, Miguel S\'anchez: {\em On smooth Cauchy hypersurfaces and Geroch's splitting theorem}. Commun.Math.Phys. 243 (2003) 461-470.  arXiv:gr-qc/0306108

\bibitem{BS2}
Antonio Bernal, Miguel S\'anchez: {\em Smoothness of time functions and the metric splitting of globally hyperbolic spacetimes}. Commun.Math.Phys. 257 (2005) 43-50.  arXiv:gr-qc/0401112 

\bibitem{BS3}
Antonio Bernal, Miguel S\'anchez: {\em Further results on the smoothability of Cauchy hypersurfaces and Cauchy time functions}. Lett.Math.Phys. 77 (2006) 183-197. arXiv:gr-qc/0512095

\bibitem{BS4}
Antonio Bernal, Miguel S\'anchez: {\em Globally hyperbolic spacetimes can be defined as "causal" instead of "strongly causal"}. Class.Quant.Grav. 24 (2007) 745-750. arXiv:gr-qc/0611138

\bibitem{CG}
Yvonne Choquet-Bruhat, Robert Geroch: {\em Global aspects of the Cauchy problem in general relativity}, Comm. Math. Phys. 14, no 4 (1969), 329-335.

\bibitem{G}
Robert Geroch: {\em Domain of dependence}. J. Math. Phys. 11 (1970), 437-449

\bibitem{MS}
Olaf M\"uller, Miguel S\'anchez: {\em Lorentzian manifolds isometrically embeddable in $L^N$}. Trans. Amer. Math. Soc. 363 (2011), 5367-5379. arXiv:0812.4439 



\end{thebibliography}
\end{document}